\documentstyle[12pt]{article}
\begin{document}
\hfill{NCKU-HEP-97-04}\par
\vskip 0.5cm
\begin{center}
{\large {\bf Unified derivation of evolution equations}}
\vskip 1.0cm
Hsiang-nan Li
\vskip 0.5cm
Department of Physics, National Cheng-Kung University, \par
Tainan, Taiwan, Republic of China
\end{center}
\vskip 1.0cm

PACS numbers: 12.38.Cy, 11.10.Hi

%\baselineskip=2\baselineskip

\centerline{\bf Abstract}
\vskip 0.3cm

We derive the evolution equations of parton distribution functions
appropriate in different kinematic regions in a unified and simple way using 
the resummation technique. They include the
Dokshitzer-Gribov-Lipatov-Altarelli-Parisi
equation for large momentum transfer $Q$, the Balitskii-Fadin-Kuraev-Lipatov 
equation for a small Bjorken variable $x$, and the 
Ciafaloni-Catani-Fiorani-Marchesini equation which embodies the above two 
equations. The relation among these equations is explored, and possible
applications of our approach are proposed.

\newpage
\centerline{\large\bf I. INTRODUCTION}
\vskip 0.5cm

Perturbative QCD (PQCD), as a gauge field theory, involves large logarithms 
from radiative corrections at each order of the coupling 
constant $\alpha_s$, such as $\ln Q$ in the kinematic region with a large 
momentum transfer $Q$ and $\ln(1/x)$ in the region with a small Bjorken 
variable $x$. These logarithms, spoiling the perturbative expansion, must be 
organized. To organize the logarithmic corrections to a parton distribution 
function, the various evolution equations have been derived. For example, 
the Dokshitzer-Gribov-Lipatov-Altarelli-Parisi (DGLAP) equation \cite{AP}
sums single logarithms $\ln Q$ for a large $x$, the 
Balitskii-Fadin-Kuraev-Lipatov (BFKL) equation \cite{BFKL} sums $\ln(1/x)$
for a small $x$, and the Ciafaloni-Catani-Fiorani-Marchesini (CCFM) equation 
\cite{CCFM}, appropriate for both large and small $x$, unifies the above two 
equations. The conventional derivation of the evolution equations usually
requires complicated diagrammatic analyses. The idea is to locate the region
of the loop momenta flowing through the rungs (radiative gluons) of a ladder 
diagram, in which leading logarithmic corrections are produced, and 
to sum the ladder diagrams to all orders. For the
DGLAP, BFKL, and CCFM equations, the important regions are those with the 
strong transverse momentum ordering, the strong rapidity ordering, and the 
strong angular ordering, respectively.

%Furthermore, it is the Reggeon
%ladder that is considered for the BFKL equation. The rung gluons, which
%obey and do not obey the angular ordering, are treated separately for the
%CCFM equation. The relation among the above equations is not clear.

In this paper we shall propose an alternative approach to the all-order 
summations of the various large logarithms. This approach is based on the 
Collins-Soper-Sterman resummation technique, which was developed originally
for the organization of double logarithms $\ln^2 Q$ \cite{CS}. Recently, we
applied this technique to some hard QCD processes, such as deep inelastic
scattering, Drell-Yan production, and inclusive heavy meson decays, and
demonstrated how to resum the double logarithms contained in parton
distribution functions into Sudakov form factors \cite{L1}. It has been
shown \cite{L1,L2} that the resummation technique is equivalent to the
Wilson-loop formalism \cite{GK,CC,FST}
for the summation of soft logarithms. Here we shall further show
that it can also deal with the single-logarithm cases, {\it i.e.}, the 
evolution equations mentioned above. Therefore, the resummation technique 
indeed has a wide application in PQCD. It will be found that our
approach is much simpler than the conventional one, and provides a unified
viewpoint to the evolution equations.

The procedures of resummation are summarized below. The 
derivative of a parton distribution function with respect to $Q$ or $x$ is 
first related to a new function, which contains a special gluon vertex. 
This relation, as a consequence of the Ward identity, is exact.
Factorizing the new function into the convolution of the subdiagram
involving the special vertex with the original parton distribution function,
we derive the evolution equation. At this stage, the kinematic orderings of
radiative gluons are specified, and the subdiagram is 
identified as the corresponding kernels.
We show explicitly that the lowest-order subdiagram gives the kernel for
the leading-logarithm summation.

In the region with both large $Q$ and small $x$ many gluons are radiated in 
scattering processes with small spatial separation among them, and a new 
effect from the annihilation of two gluons into one gluon becomes important. 
Taking into account this effect, a nonlinear evolution equation, the 
Gribov-Levin-Ryskin (GLR) equation \cite{GLR} is obtained. Using the 
resummation technique, the annihilation effect is introduced through 
next-to-leading-twist contributions to the subdiagram containing the special
vertex, and the GLR equation can be derived easily \cite{L3}. In the 
present work we shall not address this subject, because it is not very 
relevant.

We derive the DGLAP, BFKL and CCFM equations in Sect. II, III, and IV, 
respectively, by means of the resummation technique in axial gauge. We
explain how the subdiagram containing the special vertex reduces to the
evolution kernels in the different kinematic regions.
Section V is the conclusion.

\vskip 1.0cm

\centerline{\large\bf II. THE DGLAP EQUATION}
\vskip 0.5cm

Consider deep inelastic scattering (DIS) of a hadron with a light-like
momentum $p=p^+\delta^{\mu +}$ in the large $x$ limit, where 
$x=-q^2/(2p\cdot q)=Q^2/(2p\cdot q)$ is the Bjorken variable with $q$ the 
momentum transfer to the hadron through a virtual photon. According to
factorization theorems, collinear divergences arising from radiative
corrections to DIS are absorbed into a quark distribution funtion
$\phi(\xi,p^+)$ associated with the hadron, $\xi$ being the momentum
fraction of a parton. The argument $p^+$ denotes the large logarithms
$\ln p^+$ from the collinear divergences, which will be organized.

In the axial gauge $n\cdot A=A^+=0$, $n^\mu=\delta^{\mu -}$ being a 
vector on the light cone, the gauge invariant distribution function $\phi$
is defined, in the modified minimum subtraction scheme, by
\begin{equation}
\phi(\xi,p^+)=\int\frac{dy^-}{2\pi}e^{-i\xi p^+y^-}\frac{1}{2}\sum_\sigma
\langle p,\sigma| {\bar q}(y^-)\frac{1}{2}\gamma^+ q(0)|p,\sigma
\rangle|_{A^+=0}\;,
\label{dep}
\end{equation}
where $\gamma^+$ is a Dirac matrix, and $|p,\sigma\rangle$ denotes the
incoming hadron with momentum $p$ and spin $\sigma$. An average over color
is understood. To implement the resummation technique, we allow $n$ to vary
away from the light cone ({\it i.e.}, $n^2\not= 0$) temporarily, and the
resultant $n$-dependent distribution function is written as
\begin{equation}
\phi^{(n)}(\xi,p^+)=\int\frac{dy^-}{2\pi}e^{-i\xi p^+y^-}\frac{1}{2}
\sum_\sigma\langle p,\sigma| {\bar q}(y^-)\frac{1}{2}\gamma^+ q(0)|p,\sigma
\rangle|_{n\cdot A=0}\;.
\label{den}
\end{equation}
The above formula is similar to the definition for a quark decay
function employed in \cite{CS}. After deriving the evolution equation, we
let $n^\nu$ approach $\delta^{\mu -}$, and $\phi^{(n)}$ coincides with
$\phi$ in Eq.~(\ref{dep}). That is, the arbitrary vector $n$ appears only
at the intermediate stage of our formalism, and as an auxiliary tool of the
resummation technique. $\phi^{(n)}$ can be expressed as the convolution
of an infrared finite function with $\phi$, and thus they contain the same
nonperturbative information. Their relation will be investigated in details
elsewhere \cite{L5}.

The key step of resummation is to obtain the derivative
$p^+d\phi^{(n)}/dp^+$. Because of the scale invariance of $\phi^{(n)}$ in
the vector $n$ as indicated by the gluon propagator, $-iN^{\mu\nu}(l)/l^2$,
with
\begin{equation}
N^{\mu\nu}=g^{\mu\nu}-\frac{n^\mu l^\nu+n^\nu l^\mu}
{n\cdot l}+n^2\frac{l^\mu l^\nu}{(n\cdot l)^2}\;,
\label{gp}
\end{equation}
$\phi^{(n)}$ must depend on $p^+$ via the ratio $(p\cdot n)^2/n^2$. Hence,
we have the chain rule relating $p^+d/dp^+$ to $d/dn$ \cite{CS}: 
\begin{eqnarray}
p^+\frac{d}{dp^+}\phi^{(n)}=-\frac{n^2}{v\cdot n}v_\alpha\frac{d}{dn_\alpha}
\phi^{(n)}\;,
\label{cph}
\end{eqnarray}
with $v_\alpha=\delta_{\alpha +}$ a vector along $p$. The operator
$d/dn_\alpha$ applies only to the gluon propagator, giving \cite{CS}
\begin{equation}
\frac{d}{dn_\alpha}N^{\mu\nu}=
-\frac{1}{n\cdot l}(l^\mu N^{\alpha\nu}+l^\nu N^{\mu\alpha})\;.
\label{dgp}
\end{equation}
Since $p$ flows through both quark and gluon lines, while
$n$ appears only in gluon lines, the analysis is simplified by 
considering the derivative with respect to $n$, instead of $p^+$.

The loop momentum $l^\mu$ ($l^\nu$) carried by the differentiated gluon
contracts with a vertex in $\phi^{(n)}$, which is then replaced by a special
vertex \cite{CS}
\begin{eqnarray}
{\hat v}_\alpha=\frac{n^2v_\alpha}{v\cdot nn\cdot l}\;.
\label{va}
\end{eqnarray}
This special vertex can be simply read off the combination of 
Eqs.~(\ref{cph}) and (\ref{dgp}). The contraction of $l^\mu$ ($l^\nu$)
leads to the Ward identities,
\begin{equation}
\frac{i(\not k+\not l)}{(k+l)^2}(-i\not l)\frac{i\not k}{k^2}
=\frac{i\not k}{k^2}-\frac{i(\not k+\not l)}{(k+l)^2}\;,
\end{equation}
for a quark-gluon vertex, and
\begin{equation}
l^\nu\frac{-iN^{\alpha\mu}(k+l)}{(k+l)^2}\Gamma_{\mu\nu\lambda}
\frac{-iN^{\lambda\gamma}(k)}{k^2}
=-i\left[\frac{-iN^{\alpha\gamma}(k)}{k^2}-
\frac{-iN^{\alpha\gamma}(k+l)}{(k+l)^2}\right]\;,
\label{ward}
\end{equation}
for the triple-gluon vertex $\Gamma_{\mu\nu\lambda}$. The Ward identity
for the four-gluon vertex,
\begin{eqnarray}
\Gamma^{abcd}_{\lambda\mu\nu\sigma}&\propto&
f^{abe}f^{cde}(g_{\lambda\nu}g_{\mu\sigma}-g_{\lambda\sigma}g_{\mu\nu})+
f^{ace}f^{bde}(g_{\lambda\mu}g_{\nu\sigma}-g_{\lambda\sigma}g_{\mu\nu})
\nonumber \\
& &+f^{ade}f^{cbe}(g_{\lambda\nu}g_{\mu\sigma}-
g_{\lambda\mu}g_{\sigma\nu})\;,
\end{eqnarray}
is simpler. The sum of the four contractions from $l_1^{\lambda}$,
$l_2^{\mu}$, $l_3^{\nu}$, and $l_4^{\sigma}$ to the four-gluon vertex
vanishes: The contractions
from $l_1^{\lambda}$ and $l_2^{\mu}$ to the first term of
$\Gamma^{abcd}_{\lambda\mu\nu\sigma}$ cancel each other, because the
first term is antisymmetric with respect to the interchange of the indices
$\lambda$ and $\mu$. Similar cancellation occurs between the contractions
from $l_3^{\nu}$ and $l_4^{\sigma}$. The contractions from $l_1^{\lambda}$
and $l_3^{\nu}$ and from $l_2^{\mu}$ and $l_4^{\sigma}$ to the second term
cancel separately. The contractions from $l_1^{\lambda}$ and
$l_4^{\sigma}$ and from $l_2^{\mu}$ and $l_3^{\nu}$ to the third term
also cancel separately.

Summing all the diagrams
with different differentiated gluons, those embedding the special vertices
cancel by pairs, leaving the one in which the special vertex moves to the
outer end of the quark line \cite{CS}. For a fixed $x$, we obtain the
formula,
\begin{equation}
p^+\frac{d}{dp^+}\phi^{(n)}(x,p^+)=2{\bar \phi}(x,p^+)\;,
\label{dph}
\end{equation}
shown in Fig.~1(a), where $\bar \phi$, the new function mentioned in the
Introduction, contains one special vertex represented by a square. 
An identical formula to Eq.~(\ref{dph}) has been derived for a quark decay
function in \cite{CS}. The coefficient 2 comes from the equality of the
new functions with the special vertex on either of the two valence quark 
lines. Equation (\ref{dph}) is an exact consequence of the Ward 
identity without any approximation \cite{CS}. An approximation will be
introduced, when relating ${\bar \phi}$ to $\phi^{(n)}$ by factorizing out
the subdiagram containing the special vertex, such that Eq.~(\ref{dph})
becomes a differential equation of $\phi^{(n)}$.

It is known that
factorization holds only in important regions. The important regions of the
loop momentum flowing through the special vertex are soft and hard, since
the vector $n$ does not lie on the light cone, and the collinear
enhancements are suppressed. In the soft and hard regions $\bar \phi$ can
be factorized into the convolution of the subdiagram containing the special 
vertex with the original distribution function $\phi^{(n)}$,
\begin{eqnarray}
{\bar \phi}_s(x,p^+)&=&ig^2{\cal C}_F\mu^\epsilon\int
\frac{d^{4-\epsilon}l}{(2\pi)^{4-\epsilon}}
\frac{{\hat v}_\mu v_\nu}{v\cdot l}N^{\mu\nu}(l)
\left[\frac{1}{l^2}\phi^{(n)}(x,p^+)\right.
\nonumber \\
& &\left.+2\pi i\delta(l^2)\phi^{(n)}(x+l^+/p^+,p^+)\right]
-\delta K\phi^{(n)}(x,p^+)\;,
\label{kj}\\
{\bar \phi}_h(x,p^+)&=&-ig^2{\cal C}_F\mu^\epsilon
\int\frac{d^{4-\epsilon}l}{(2\pi)^{4-\epsilon}}{\hat v}_\mu
\frac{N^{\mu\nu}(l)}{l^2}\left[\frac{\xi\not p-\not l}{(\xi p-l)^2}\gamma_\nu
+\frac{v_\nu}{v\cdot l}\right]\phi^{(n)}(x,p^+)
\nonumber \\
& &-\delta G\phi^{(n)}(x,p^+)\;,
\label{gph}
\end{eqnarray}
where ${\cal C}_F=4/3$ is the color factor, and $\delta K$ and $\delta G$
are additive counterterms. The function ${\bar \phi}_s$, absorbing the
soft divergences of the subdiagram, corresponds to Fig.~1(b), where the
eikonal approximation for the valence quark propagator has been made.
The eikonal propagator is represented by a double line in the figure. The
first and second terms in the integral are associated with the virtual and
real gluon emissions, respectively, where $\phi^{(n)}(x+l^+/p^+,p^+)$
implies that the parton coming out of the hadron carries the longitudinal
momentum $xp^++l^+$ in order to radiate a real gluon of momentum $l^+$.
The function ${\bar \phi}_h$,
absorbing the ultraviolet divergences, corresponds to Fig.~1(c), where the
subtraction of the second diagram ensures that the involved loop momentum
is hard. We emphasize that the factorization formulas in Eqs.~(\ref{kj}) and
(\ref{gph}) are not exact, but hold only up to the leading logarithms
$\ln p^+$.

Employing the variable change $\xi=x+l^+/p^+$, and performing the
integrations in Eqs.~(\ref{kj}) and (\ref{gph}) straightforwardly, we
arrive at
\begin{eqnarray}
{\bar \phi}(x,p^+)&=&{\bar \phi}_s(x,p^+)+{\bar \phi}_h(x,p^+)
\nonumber \\
&=&\int_x^1 d\xi \left[K(x,\xi,p^+,\mu)+G(x,\xi,p^+,\mu)\right]
\phi^{(n)}(\xi,p^+)\;,
\label{con}
\end{eqnarray}
with
\begin{eqnarray}
K&=&\frac{\alpha_s(\mu)}{\pi\xi}{\cal C}_F\left[\frac{1}{(1-x/\xi)_+}
+\ln\frac{\nu p^+}{\mu}\delta(1-x/\xi)\right]\;,
\nonumber\\
G&=&-\frac{\alpha_s(\mu)}{\pi\xi}{\cal C}_F\ln\frac{\xi\nu p^+}{\mu}
\delta(1-x/\xi)\;,
\label{kgir}
\end{eqnarray}
where constants of order unity have been dropped, and
$\nu=\sqrt{(v\cdot n)^2/|n^2|}$ is the gauge factor. Equation (\ref{kgir})
confirms our argument that $\phi^{(n)}$ depends on $p$ and $n$ through the
ratio $(p\cdot n)^2/n^2=(\nu p^+)^2$. In the region with $x\to 1$ 
the logarithm $\ln(\xi\nu p^+/\mu)$ in $G$ can be replaced by
$\ln(\nu p^+/\mu)$.

We then treat $K$ and $G$ by renormalization group (RG) methods:
\begin{equation}
\mu\frac{d}{d\mu}K=-\lambda_K=-\mu\frac{d}{d\mu}G\;.
\label{kg}
\end{equation}
The anomalous dimension of $K$ is defined by $\lambda_K=-\mu d\delta K
/d\mu$, whose explicit expression is not essential here. When solving 
Eq.~(\ref{kg}), we allow the variable $\mu$ to evolve from the scale of $K$ 
to the scale of $G$. The RG solution of $K+G$ is given by
\begin{eqnarray}
K(x,\xi,p^+,\mu)+G(x,\xi,p^+,\mu)&=&K(x,\xi,p^+,p^+)+G(x,\xi,p^+,p^+)
\nonumber \\
& &-\int_{p^+}^{p^+}\frac{d{\bar\mu}}{\bar\mu}
\lambda_K(\alpha_s({\bar\mu}))\;,
\nonumber \\
&=&\frac{\alpha_s(p^+)}{\pi\xi}{\cal C}_F\frac{1}{(1-x/\xi)_+}\;,
\label{skg}
\end{eqnarray}
where the initial conditions $K(x,\xi,p^+,p^+)$ and $G(x,\xi,p^+,p^+)$ do 
not contain large logarithms after choosing $\mu$ as $p^+$. Note
that the gauge factor $\nu$ cancels between $K$ and $G$, implying the
gauge invariance of the evolution kernel $K+G$.

A remark is in order. The source of double logarithms, {\it i.e.}, the
integral containing $\lambda_K$, vanishes as shown in Eq.~(\ref{skg}). 
If the transverse degrees of freedom of a parton are
included, an extra factor $\exp(i{\bf l}_T\cdot {\bf b})$, $b$ being the
conjugate variable of the transverse momentum $k_T$ carried by the parton,
will be associated with the real gluon emission \cite{CS,L1}. Then we deal
with a two-scale problem. Approximating
$\phi^{(n)}(x+l^+/p^+,k_T,p^+)$ by $\phi^{(n)}(x,k_T,p^+)$,
the integration over $l^+$ can also be performed, and the convolution in
Eq.~(\ref{con}) is simplified to a multiplication in the $b$ space,
\begin{eqnarray}
{\bar \phi}(x,b,p^+)=\left[K(x,b,\mu)+G(x,p^+,\mu)\right]
\phi^{(n)}(x,b,p^+)\;,
\label{apm}
\end{eqnarray}
with \cite{L1} 
\begin{eqnarray}
K(x,b,\mu)=\frac{\alpha_s(\mu)}{\pi}{\cal C}_F
\ln\frac{1}{b\mu}\;.
\end{eqnarray}
Hence, $K$ is characterized by the smaller scale $1/b$, and
$\mu$ should be set to $1/b$ in order to minimize the large logarithms.
Consequently, $1/b$ is substituted for the lower bound of $\bar\mu$ in
Eq.~(\ref{skg}), and double logarithms exist. This is the difference
between this work and \cite{CS}. The above discussion indicates
that the resummation technique is applicable to the single-logarithm
as well as double-logarithm cases.

Inserting Eq.~(\ref{skg}) into (\ref{con}), Eq.~(\ref{dph}) becomes
\begin{eqnarray}
p^+\frac{d}{dp^+}\phi^{(n)}(x,p^+)=\frac{\alpha_s(p^+)}{\pi}
\int_x^1 \frac{d\xi}{\xi} P(x/\xi)\phi^{(n)}(\xi,p^+)\;,
\label{sp3}
\end{eqnarray}
with the kernel
\begin{eqnarray}
P(z)={\cal C}_F\frac{2}{(1-z)_+}\;.
\end{eqnarray}
Now we make $n^\mu$ approach $\delta^{\mu -}$ (the light cone), and obtain
\begin{eqnarray}
Q\frac{d}{dQ}\phi(x,Q)=\frac{\alpha_s(Q)}{\pi}
\int_x^1 \frac{d\xi}{\xi} P(x/\xi)\phi(\xi,Q)\;,
\label{sp2}
\end{eqnarray}
where the variable $p^+$ has been replaced by $Q$, and the gauge invariance
of the distribution function is restored. It is easy to identify $P$ as the
splitting function $P_{qq}$ in the limit $z\to 1$,
\begin{eqnarray}
P_{qq}(z)={\cal C}_F\frac{1+z^2}{(1-z)_+}\;.
\end{eqnarray}
Hence, Eq.~(\ref{dph}) leads to the DGLAP equation (\ref{sp2}), and the
diagrams in Fig.~1(b) give the splitting function.

Note that only the term of $P_{qq}$ singular at $z\to 1$, {\it i.e.},
$\xi\to x$, is reproduced in the resummation approach. The reason is
as follows. The $z^2$ term in the numerator of $P_{qq}$ arises from
the radiative correction with the ends of a real gluon attaching each
valence quark line, whose loop integral is written as
\begin{eqnarray}
I&=&g^2{\cal C}_F\int\frac{d^4l}{(2\pi)^4}
Tr\left[\gamma_\mu\frac{\xi\not p-\not l}{(\xi p-l)^2}
\frac{\gamma^+}{2}\frac{\xi\not p-\not l}{(\xi p-l)^2}\gamma_\nu
\not p\right]
\nonumber\\
& &\times N^{\mu\nu}(l)2\pi\delta(l^2)
\delta\left(\xi-x-\frac{l^+}{p^+}\right)\;.
\label{z2}
\end{eqnarray}
$\not p$ in the trace comes from the Dirac structure of the proton
distribution function. The numerator $\xi\not p-\not l$ contributes the
factor $\xi p^+ - l^+=xp^+$ because of the last $\delta$ function. The
denominator $(\xi p-l)^2=-2\xi p^+v\cdot l$ gives the factor $\xi p^+$ due
to the on-shell gluon with $l^2=0$. Their combination leads to a power of
$z=x/\xi$. The two fermion propagators then explain the term $z^2$.

When applying the derivative in Eq.~(\ref{cph}) to $N^{\mu\nu}$, we
have Eq.~(\ref{dgp}), whose first term renders Eq.~(\ref{z2})
reduce to
\begin{eqnarray}
I&=&-g^2{\cal C}_F\int\frac{d^4l}{(2\pi)^4}
Tr\left[\frac{\gamma^+}{2}\frac{\xi\not p-\not l}{(\xi p-l)^2}\gamma_\nu
\not p\right]
\nonumber\\
& &\times {\hat v}_\alpha N^{\alpha\nu}(l)2\pi\delta(l^2)
\delta\left(\xi-x-\frac{l^+}{p^+}\right)\;,
\label{z21}
\end{eqnarray}
as described by Fig.~2. To obtain the above expression, we have used
\begin{equation}
\not l\frac{\xi\not p-\not l}{(\xi p-l)^2}=
\frac{\xi\not l\not p}{-2\xi p^+v\cdot l}=
-1+\frac{\xi\not p\not l}{2\xi p^+v\cdot l}\;,
\end{equation}
where the second term gives a vanishing
contribution when multiplied by $\not p$ in the trace. Hence, the
differentiation with respect to $n$ employed in the resummation technique
removes a power of $z$. If further applying the eikonal approximation to
the remaining fermion propagator,
\begin{equation}
\frac{\xi\not p-\not l}{(\xi p-l)^2}\gamma_\nu\not p\approx
\frac{\xi\not p}{-2\xi p^+v\cdot l}\gamma_\nu\not p=
\frac{2\xi p^+v_\nu}{-2\xi p^+v\cdot l}\not p=
-\frac{v_\nu}{v\cdot l}\not p\;,
\end{equation}
Fig.~2 reduces to the seccond diagram in Fig.~1(b), and another power of
$z$ disappears.

We emphasize that the resummation technique is based on PQCD factorization
theorems, and the factorization of the subdiagram makes sense only in the 
soft and hard regions. That is, the function ${\bar \phi}_s$, with the
eikonal approximation, collects the leading contribution from the soft
region $\xi\to x$, and ${\bar \phi}_h$ collects the leading contribution
from the hard region. The subdiagram in Fig.~2, containing a finite piece
of the splitting function, can not be absorbed into ${\bar \phi}_s$
or ${\bar \phi}_h$, since the quark propagator is not eikonalized, and
the real gluon is not hard. Therefore, the finite part of the splitting
function is completely missing in the resummation approach. In the
conventional derivation of the DGLAP equation one-loop diagrams are
computed explicitly without resort to eikonal (soft) approximation, and
this makes the difference. The same conclusion applies to other cases that
involve a soft approximation, such as the conventional derivation of the
CCFM equation \cite{CCFM}, where only the terms of the relevant splitting
function singular at $z\to 1$ and at $z\to 0$ were reproduced, and the
finite part was in fact put in by hand, as mentioned in Sect. IV.

\vskip 1.0cm

\centerline{\large\bf III. THE BFKL EQUATION}
\vskip 0.5cm

In this section we demonstrate that the resummation technique reduces to the 
BFKL equation for the gluon distribution function in the small $x$ region. 
The unintegrated gluon distribution function $F(x,k_T)$, defined by
\begin{eqnarray}
F(x,k_T)&=&\frac{1}{p^+}\int\frac{dy^-}{2\pi}\int\frac{d^2y_T}{4\pi}
e^{-i(xp^+y^--{\bf k}_T\cdot {\bf y}_T)}
\nonumber \\
& &\times\frac{1}{2}\sum_\sigma
\langle p,\sigma| F^+_\mu(y^-,y_T)F^{\mu+}(0)|p,\sigma\rangle\;,
\label{deg}
\end{eqnarray}
in the axial gauge $n\cdot A=0$, describes the probability of a gluon
carrying a longitudinal momentum fraction $x$ and transverse
momenta ${\bf k}_T$. $F^+_\mu$ is the field tensor. Similarly, we vary
the vector $n$ arbitrarily first to work out the resummation, and
then show that the BFKL kernel is independent of $n$ as in the DGLAP
case. After deriving the evolution equation, $n$ is brought back to the
light cone.

We do not make explicit the $p^+$ dependence of $F$ for the following
reason. The BFKL equation governs the behavior of $F$ with the momentum
fraction $x$. The variation of $x$ can be achieved by varying the hadron
momentum $p^+$, if the parton momentum $xp^+=k^+$, appearing only in the
exponent in Eq.~(\ref{deg}), is fixed. Again, $F$ involves the vectors
$k$ and $n$, which should combine into the ratio $(k\cdot n)^2/n^2$.
Since $k^+$ is fixed, we do not choose it as an argument of $F$, and regard
that $F$ depends on $p^+$ implicitly through $x=k^+/p^+$.
Hence, the derivative of $F$ with respect to $x$ is related to the
derivative with respect to $p^+$ considered in the resummation formalism,
and the chain rule in Eq.~(\ref{cph}) holds:
\begin{equation}
-x\frac{d}{dx}F(x,k_T)=p^+\frac{d}{dp^+}F(x,k_T)=
-\frac{n^2}{v\cdot n}v_\alpha\frac{d}{dn_\alpha}F(x,k_T)\;.
\end{equation}
Applying the operator $d/dn$ to a gluon propagator,
we get Eq.~(\ref{dgp}) and the same special vertex in Eq.~(\ref{va}).
Summing all the diagrams with different differentiated gluons and employing
the Ward identity,
the special vertex moves to the outer end of the parton line as explained
in Sect. II \cite{CS}. We then obtain the derivative of $F$,
\begin{equation}
-x\frac{d}{dx}F(x,p_T)=2{\bar F}(x,p_T)\;,
\label{df}
\end{equation}
described by Fig.~3(a), where the new function $\bar F$ contains one
special vertex. It is easy to observe that Eq.~(\ref{df}) is the copy of
Eq.~(\ref{dph}) for the gluon distribution function.

We relate ${\bar F}$ to $F$ by factorizing out the subdiagram
containing the special vertex, such that Eq.~(\ref{df}) reduces to a
differential equation of $F$. In the leading soft and hard regions of
$l$, which flows through the special vertex, the factorization is performed
according to Figs.~3(b) and 3(c), respectively. Fig.~3(b) collects the soft
divergences by eikonalizing the gluon propagator.
We extract the color factor from the relation
$f_{abc}f_{bdc}=-N_c\delta_{ad}$, where the indices $a,b,\dots$ have been
indicated in the figure, and $N_c=3$ is the number of colors. The
corresponding factorization formula is written as
\begin{eqnarray}
{\bar F}_s(x,k_T)&=&
iN_cg^2\int\frac{d^{4}l}{(2\pi)^4}
\Gamma_{\mu\nu\lambda}{\hat v}_\beta
[-iN^{\nu\beta}(l)]
\frac{-iN^{\lambda\gamma}(xp)}{-2xp\cdot l}
\nonumber \\
& &\times\left[2\pi i\delta(l^2)F(x,|{\bf k}_T+{\bf l}_T|)
+\frac{\theta(k_T^2-l_T^2)}{l^2}F(x,k_T)\right]\;,
\label{kf}
\end{eqnarray}
where $iN_c$ comes from the product of the overall coefficient $-i$ in
Eq.~(\ref{ward}) and the color factor $-N_c$ extracted above, 
and the triple-gluon vertex for vanishing $l$ is given by
\begin{equation}
\Gamma_{\mu\nu\lambda}=
-g_{\mu\nu}xp_{\lambda}-g_{\nu\lambda}xp_{\mu}+2g_{\lambda\mu}xp_{\nu}\;.
\label{tri}
\end{equation}
The first term in the brackets
corresponds to the real gluon emission, where
$F(x,|{\bf k}_T+{\bf l}_T|)$ implies that the parton coming out of
the hadron carries the transverse momenta 
${\bf k}_T+{\bf l}_T$ in order to radiate a real gluon of momenta
${\bf l}_T$. The second term corresponds to the virtual gluon emission,
where the $\theta$ function sets the upper bound of $l_T$ to $k_T$ to ensure
a soft momentum flow.

It can be shown that the contraction of $p$ with a vertex in the
quark box diagram the partons attach, or with a vertex in the gluon
distribution function, leads to a
contribution down by a power $1/s$, $s=(p+q)^2$, compared to the
contribution from the contraction with ${\hat v}_\beta$. Following this
observation, Eq.~(\ref{kf}) is expressed as 
\begin{eqnarray}
{\bar F}_s(x,k_T)&=&
iN_cg^2\int\frac{d^{4}l}{(2\pi)^4}N^{\nu\beta}(l)
\frac{{\hat v}_\beta v_\nu}{v\cdot l}
\left[2\pi i\delta(l^2)F(x,|{\bf k}_T+{\bf l}_T|)\right.
\nonumber \\
& &\left.
+\frac{\theta(k_T^2-l_T^2)}{l^2}F(x,k_T)\right]\;,
\label{kf1}
\end{eqnarray}
which corresponds to Fig.~3(b) exactly.
The eikonal vertex $v_\nu$ comes from the last term divided by
$xp^+$ in Eq.~(\ref{tri}).
The remaining metric tensor $g^{\mu\gamma}$ has been absorbed into $F$.
Assuming $n=(n^+,n^-,{\bf 0})$ for convenience,
the integrations over $l^-$ and $l^+$ to infinity in Eq.~(\ref{kf1}) give
\begin{eqnarray}
{\bar F}_s(x,k_T)&=&\frac{{\bar \alpha}_s}{2}
\int\frac{d^{2}l_T}{\pi}
\frac{-n^2}{2n^-[n^+l_T^2+2n^-l^{+2}]}|^{l^+=\infty}_{l^+=0}
\nonumber\\
& &\times\left[F(x,|{\bf k}_T+{\bf l}_T|)
-\theta(k_T^2-l_T^2)F(x,k_T)\right]\;,
\nonumber\\
&=&\frac{{\bar \alpha}_s}{2}
\int\frac{d^{2}l_T}{\pi l_T^2}
\left[F(x,|{\bf k}_T+{\bf l}_T|)-\theta(k_T^2-l_T^2)F(x,k_T)\right]\;,
\label{kf2}
\end{eqnarray}
with ${\bar \alpha}_s=N_c\alpha_s/\pi$. The first line of the above
formulas demonstrates explicitly how the $n$ dependnece cancels in the
evaluation of Fig.~3(b).

The contribution from the first diagram of Fig.~3(c) is written as
\begin{eqnarray}
& &\frac{{\bar\alpha}_s}{2}\int\frac{d^{2}l_T}{\pi}
\left[\frac{1}{l_T^2}-\frac{1}{l_T^2+(k^+\nu)^2}\right.
\nonumber \\
& &\left.
-\frac{1}{2}\frac{k^+\nu}{[l_T^2+(k^+\nu)^2]^{3/2}}
\ln\frac{\sqrt{l_T^2+(k^+\nu)^2}-k^+\nu}
{\sqrt{l_T^2+(k^+\nu)^2}+k^+\nu}\right]\;,
\label{g1}
\end{eqnarray}
which confirms the statement that $F$ depends on $x$ (or $p^+$) via the
ratio $(k\cdot n)^2/n^2=(k^+\nu)^2$.
In the interesting region with small $k^+$, Eq.~(\ref{g1}) is less
important (it vanishes as $k^+\to 0$), and ${\bar F}_s$ dominates.
Ignoring the contribution from Fig.~3(c) along with its soft subtraction
(the second diagram), that is, adopting ${\bar F}\approx {\bar F}_s$,
Eq.~(\ref{df}) becomes
\begin{eqnarray}
\frac{dF(x,k_T)}{d\ln(1/x)}=
{\bar \alpha}_s\int\frac{d^{2}l_T}{\pi l_T^2}
\left[F(x,|{\bf k}_T+{\bf l}_T|)-\theta(k_T^2-l_T^2)F(x,k_T)\right]\;,
\label{bfkl}
\end{eqnarray}
which is exactly the BFKL equation. The $n$ dependence residing in Fig.~3(c)
is removed with the vanishing of Eq.~(\ref{g1}) at small $k^+$, and the BFKL
kernel turns out to be gauge invariant. It is understood that the
diagrams in Fig.~3(b) play the role of the BFKL kernel, a similar
conclusion to that drawn at the end of Sect. II.

Since the explicit dependence of $F$ on the large scale $p^+$ is neglected,
the transverse degrees of freedom of a parton must be
taken into account. This explains why the gluon distribution function at
small $x$ is constructed based on the high-energy
$k_T$-factorization theorem \cite{J}. Therefore,
our formalism is applicable to the distribution functions
defined according to the collinear factorization (the DGLAP case) and 
according to the $k_T$-factorization (the BFKL case). 

\vskip 1.0cm

\centerline{\large\bf IV. THE CCFM EQUATION}
\vskip 0.5cm

With the discussion in the previous two sections, it is not difficult
to demonstrate that the resummation technique reduces to the CCFM equation,
which embodies the DGLAP equation and the BFKL equation. It hints that we
should maintain both the $l^+$ and $l_T$ dependences in the unintegrated 
distribution function for the real gluon emission, namely, consider
$F(x+l^+/p^+,|{\bf k}_T+{\bf l}_T|)$.
The BFKL equation is appropriate for the multi-Regge region,
where the transverse momenta carried by the rung gluons of a ladder diagram 
are of the same order, {\it i.e.}, $l_T\approx k_T$. Hence, the loop
momentum $l_T$ flowing through the distribution function is not
negligible, and we can make the soft approximation 
\begin{equation}
F(x+l^+/p^+,|{\bf k}_T+{\bf l}_T|)\approx
F(x,|{\bf k}_T+{\bf l}_T|)\;,
\label{ff}
\end{equation}
in ${\bar F}_s$, from which the BFKL equation is 
derived. While the DGLAP equation is appropriate for the transverse
momemtum ordered region, in which we have $l_T\ll p_T$, and thus the
approximation
\begin{equation}
F(x+l^+/p^+,|{\bf k}_T+{\bf l}_T|)\approx F(x+l^+/p^+,k_T)\;.
\end{equation}
The $k_T$ dependence of
the distribution function decouples, and is integrated out from
both sides of Eq.~(\ref{bfkl}). This is the reason
a parton distribution function in the DGLAP equation needs not to
involve the transverse degrees of freedom. The argument $x+l^+/p^+$ then
leads to the splitting function. 
The same diagrams in Figs.~1(b) and 3(b) give the different evolution
kernels, because they are factorized according to the different kinematic
orderings of radiative gluons.

Start with
\begin{equation}
p^+\frac{d}{dp^+}F(x,k_T,p^+)=2{\bar F}(x,k_T,p^+)\;,
\label{cc}
\end{equation}
where the arguments $k_T$ and $p^+$ of the unintegrated gluon distribution
function manifest the attempt to unify the BFKL and DGLAP equations.
Similarly, the new function
$\bar F$ involves one special vertex at the outer end of a parton line.
If following the standard procedures of resummation, we should factorize out
the subdiagram containing the special vertex in the leading soft and hard
regions, and derive ${\bar F}_s$ and ${\bar F}_h$, respectively, as in
the previous sections. The function ${\bar F}_h$ involves the lowest-order
virtual gluon emission. This idea leads to a new unified evolution equation,
which will be studied elsewhere \cite{LL}. To reproduce the CCFM equation,
however, the inclusion of virtual gluons must be performed in a different
way: They are embedded in ${\bar F}_s$, instead of absorbed into
${\bar F}_h$. Consequently, the factorization of the subdiagram is
described by Fig.~4(a), where the two jet functions $J$ group
all-order virtual corrections, and the lowest-order real gluon emission
between them is soft.

First, we resum the double logarithms contained in $J$ by considering its
derivative
\begin{eqnarray}
p^+\frac{d}{dp^+}J(p_T,p^+)&=&{\bar J}(k_T,p^+)
\nonumber \\
&=&[K_J(k_T,\mu)+G_J(p^+,\mu)]J(k_T,p^+)\;,
\label{ccj}
\end{eqnarray}
which is similar to Eq.~(\ref{apm}). At lowest order the function $K_J$
comes from the first diagram of Fig.~3(b), and $G_J$ from the two diagrams
in Fig.~3(c). The relation between $K_J+G_J$
and $J$ is simply multiplicative, since $J$ collects only virtual gluons. 
We have set the infrared cutoff of $K_J$ to $k_T$, as indicated by its 
argument. This cutoff is necessary here due to the lack of the corresponding 
real gluon emission, which serves as a soft regulator. The one-loop $K_J$ 
can be obtained simply by working out the second integral in 
Eq.~(\ref{kf1}) without the $\theta$ function. The anomalous dimension of 
$K_J$ is then found to be $\gamma_J={\bar\alpha_s}$. The function $G_J$ 
can also be computed, but its explicit expression is not important.
The standard RG analysis leads to
\begin{equation}
K_J(k_T,\mu)+G_J(p^+,\mu)=
-\int_{k_T}^{p^+}\frac{d{\bar\mu}}{\bar\mu}\gamma_J(\alpha_s({\bar\mu}))\;,
\label{cckg}
\end{equation}
with the initial conditions $K_J(k_T,k_T)=G_J(p^+,p^+)=0$. Of course, we
have neglected the constants of order unity in $K_J$ and $G_J$.

Substituting Eq.~(\ref{cckg}) into (\ref{ccj}), we solve for 
\begin{equation}
J(k_T,Q)=\Delta(Q,k_T)J^{(0)}\;,
\label{jd}
\end{equation}
with the double-logarithm exponential 
\begin{eqnarray}
\Delta(Q,k_T)=\exp\left[-{\bar\alpha_s}
\int_{k_T}^{Q}\frac{dp^+}{p^+}
\int_{k_T}^{p^+}\frac{d{\bar\mu}}{\bar\mu}\right]\;.
\end{eqnarray}
We have chosen the upper bound of $p^+$ as $Q$, and ignored the running of
${\bar\alpha}_s$. The initial condition $J^{(0)}$ can be regarded as a
tree-level gluon propagator, and then eikonalized in the evaluation of the
soft real gluon emission below. We split the above exponential into
\begin{equation}
\Delta(Q,k_T)=
\Delta_S^{1/2}(Q,zq)\Delta_{NS}^{1/2}(z,q,k_T)\;, 
\end{equation}
with $z=x/\xi$ and $q=l_T/(1-z)$, where $\xi$ is the momentum fraction
entering $J$ from the bottom, and $l_T$ is the transverse loop momentum 
carried by the real gluon. The so-called ``Sudakov" exponential
$\Delta_S$ and the ``non-Sudakov" exponential $\Delta_{NS}$ are given by
\begin{eqnarray}
\Delta_S(Q,zq)&=&\exp\left[-2{\bar\alpha_s}
\int_{zq}^{Q}\frac{dp^+}{p^+}
\int_{k_T}^{p^+}\frac{d{\bar\mu}}{\bar\mu}\right]
\nonumber \\
&=&\exp\left[-{\bar\alpha_s}
\int_{(zq)^2}^{Q^2}\frac{dp^2}{p^2}
\int_{0}^{1-k_T/p}\frac{dz'}{1-z'}\right]
\nonumber \\
\Delta_{NS}(z,q,p_T)&=&\exp\left[-2{\bar\alpha_s}
\int_{k_T}^{zq}\frac{dp^+}{p^+}
\int_{k_T}^{p^+}\frac{d{\bar\mu}}{\bar\mu}\right]\;.
\nonumber \\
&=&\exp\left[-{\bar\alpha_s}\int_{z}^{k_T/q}\frac{dz'}{z'}
\int_{(z'q)^2}^{k_T^2}\frac{dp^2}{p^2}\right]\;,
\label{nons}
\end{eqnarray}
where the variable changes ${\bar \mu}=(1-z')p$ and $p^+=p$ for $\Delta_S$, 
and ${\bar \mu}=p$ and $p^+=z'q$ for $\Delta_{NS}$ have been employed. 

With Eq.~(\ref{jd}), Fig.~4(a) reduces to Fig.~4(b), where the tree-level
gluon propagator $J^{(0)}$ on the right-hand side has been eikonalized, and
that of the left-hand side has been absorbed into $F$. 
Based on Fig.~4(b), $\bar F$ is written as
\begin{eqnarray}
{\bar F}(x,k_T,p^+)&=&iN_cg^2\int\frac{d^4l}{(2\pi)^4}
N^{\nu\beta}(l)\frac{{\hat v}_\beta v_\nu}{v\cdot l}
2\pi i\delta(l^2)\Delta^2(Q,k_T)
\nonumber \\
& &\times \theta(Q-zq)
F(x+l^+/p^+,|{\bf k}_T+{\bf l}_T|,p^+)\;.
\label{ci}
\end{eqnarray}
Basically, the above formula is similar to the
real gluon emission term in Eq.~(\ref{kf1}) except for the exponential
$\Delta^2$ from the two jet functions $J$, and the $\theta$ function. The
$\theta$ function requires $Q>zq$ such that the Sudakov exponential
$\Delta_S$ is meaningful, which comes from the angular ordering of 
radiative gluons $Q/(xp^+) > l_T/[(\xi-x)p^+]$. Compared with the
transverse momentum ordering for the DGLAP equation \cite{CCFM,KMS}, $Q$
($l_T$) has been divided by the gluon energy $xp^+$ ($(\xi-x)p^+$). Hence,
the inserted scale $zq$ reflects the special kinematic ordering for the
CCFM equation. Those radiative gluons, which do not obey the angular
ordering, conntribute to the non-Sudakov exponential. This is one of the
motivations to introduce the scale $zq$.

Using the variable change
$\xi=x+l^+/p^+$ and performing the integration over $l^-$, we obtain
\begin{eqnarray}
{\bar F}(x,k_T,p^+)&=&\frac{\bar\alpha_s}{2}\int_x^1 d\xi
\int\frac{d^2l_T}{\pi}
\frac{2n^2(\xi-x)p^{+2}}{[n^+l_T^2+2n^-(\xi-x)^2p^{+2}]^2}\Delta^2(Q,k_T)
\nonumber\\ 
& &\times \theta(Q-zq)F(\xi,|{\bf k}_T+{\bf l}_T|,p^+)\;,
\label{cctf}
\end{eqnarray}
where $n=(n^+,n^-,{\bf 0})$ has been assumed.
Equation (\ref{cctf}) is then substituted into (\ref{cc}) to find the
solution of $F$. We integrate Eq.~(\ref{cctf}) from $p^+=0$ to $Q$,
and adopt the variable changes $\xi=x/z$ and ${\bf l}_T=(1-z){\bf q}$.
To work out the $p^+$ integration, $F(x/z,|{\bf k}_T+{\bf l}_T|,p^+)$
is approximated by $F(x/z,|{\bf p}_T+{\bf l}_T|,l_T)$.
The solution of $F$ is given by
\begin{eqnarray}
F(x,k_T,Q)&=&F^{(0)}+{\bar\alpha_s}\int_x^1 dz
\int\frac{d^2q}{\pi q^2}\theta(Q-zq)
\Delta_S(Q,zq)\Delta_{NS}(z,q,k_T)
\nonumber \\
& &\hspace{2.0cm}\times
\frac{1}{z(1-z)}F(x/z,|{\bf k}_T+(1-z){\bf q}|,l_T)\;,
\label{ccc1}
\end{eqnarray}
where the nonperturbative initial condition $F^{(0)}$ corresponds to the 
lower bound of $p^+$. Again, the $n$ dependence disappears for a
similar reason to that for Eq.~(\ref{kf2}), and the CCFM kernel is
gauge invariant.

Equation (\ref{ccc1}) can be reexpressed as
\begin{eqnarray}
F(x,k_T,Q)&=&F^{(0)}+\int_x^1 dz
\int\frac{d^2q}{\pi q^2}\theta(Q-zq)
\Delta_S(Q,zq){\tilde P}(z,q,k_T)
\nonumber \\ 
& &\hspace{2.0cm}\times
F(x/z,|{\bf k}_T+(1-z){\bf q}|,l_T)\;,
\label{ccfm}
\end{eqnarray}
with 
\begin{equation}
{\tilde P}={\bar\alpha_s}
\left[\frac{1}{1-z}+\Delta_{NS}(z,q,k_T)\frac{1}{z}-2+z(1-z)\right]\;,
\label{tp}
\end{equation}
which is close to the splitting function
\begin{equation}
P_{gg}={\bar\alpha_s}
\left[\frac{1}{1-z}+\frac{1}{z}-2+z(1-z)\right]\;.
\label{pgg}
\end{equation}
Obviously, Eq.~(\ref{ccfm}) is the CCFM equation \cite{CCFM}. 
To arrive at Eq.~(\ref{tp}), we have employed the identity $1/(z(1-z))
\equiv 1/(1-z) +1/z$, and put in by hand the last term $-2+z(1-z)$. This
term, finite at $z\to 0$ and at $z\to 1$, can not
be obtained in the conventional approach either \cite{CCFM} as stated at the 
end of Sect. II. Note that only the non-Sudakov form factor $\Delta_{NS}$ 
in front of $1/z$ is kept, because $\Delta_{NS}$ vanishes when the upper 
bound $zq$ of $p^+$ approaches zero, as shown in Eq.~(\ref{nons}), and thus 
smears the $z\to 0$ pole of the function ${\tilde P}$.
This is another motivation to introduce the scale $zq$. The above
derivation shows that the complicated diagrammatic analysis
involved in the conventional derivation of the CCFM equation \cite{CCFM}
is avoided using the resummation technique.

\vskip 1.0cm

\centerline{\large\bf VI. CONCLUSION}
\vskip 0.5cm

In this paper we have shown that the resummation technique provides a 
unified and simple viewpoint to the organization of the various large
logarithms, and reduces to the DGLAP equation, the BFKL equation, and the 
CCFM equation in the different kinematic regions. The main idea is to relate
the derivative of a parton distribution function to a new function involving
a special vertex. The summation of the large logarithms is embedded in 
the new function without resort to complicated diagrammatic analyses. 
When expressing the new function as a factorization formula, we obtain the 
evolution equation. The subdiagram containing the special vertex,
factorized according to the specific orderings of radiative gluons, leads
to the corresponding kernels. By means of the resummation technique, the
connection among the evolution equations becomes transparent.

Many applications of our formalism to small $x$ physics follow this work,
which will be briefly described below. In Eq.~(\ref{kf}) associated with
the BFKL equation, the real gluon emission term in fact involves the
distribution function $F(x+l^+/p^+,|{\bf k}_T+{\bf l}_T|)$ as in the
DGLAP and CCFM cases, instead of $F(x,|{\bf k}_T+{\bf l}_T|)$. The former
is replaced by the latter when the strong rapidity ordering
$x+l^+/p^+ \gg x$ {\it i.e.}, Eq.~(\ref{ff}) is employed. Taking this into
account, it is of no doubt that the loop momentum $l^+$ should not be
extended to infinity in the derivation of Eq.~(\ref{kf2}), since
$F(x+l^+/p^+)$ vanishes at large momentum fraction. If truncating $l^+$ at
the scale of order $Q$, the resultant BFKL equation contains an intrinsic
$Q$ dependence. This modified equation has been proposed and studied in
\cite{L4}. The HERA data of the DIS structure function $F_2(x,Q^2)$
\cite{H1}, which exhibit a steep rise (corresponding to hard pomeron
exchanges) at small $x$ for large $Q^2\sim 50$ GeV$^2$, and a flat
rise for low $Q^2\sim 4$ GeV$^2$ (because soft pomeron contributionns
begin to play), were explained successfully. Note that the $Q$ dependence
of the data can not be understood by means of the conventional BFKL
equation.

Another consequence of relaxing the rapidity ordering is the recovery of
the unitarity of the BFKL evolution. It can be easily recognized that real
gluon emissions are responsible for the rise of the gluon distribution
and the structure function $F_2$. However, the approximation in
Eq.~(\ref{ff}) overestimates the real gluon contributions, such that
$F_2$ rises as a power $x^{-\lambda}$, and violates the unitarity bound
$F_2 \le {\rm const}. \ln^2(1/x)$. By adopting $F(x+l^+/p^+)$ in the
evaluation of the BFKL kernel, we have been able to show that the power
rise is moderated into a logarithmic rise \cite{L6}. The reason we have
more freedom to modify the BFKL equation is that the gluons are not
reggeized according to the rapidity ordering before the BFKL kernel is
derived.

The BFKL equation for the polarized gluon distribution function can
be derived simply in our formalism \cite{LY}. After factorizing the
subdiagram containing the special vertex, we contract Eq.~(\ref{kf})
with the gluon polarization vectors $\epsilon_\mu \epsilon_\gamma$
for positive and negative helicities, and take their difference.
The obtained equation will be employed to study the small $x$ behavior
of the polarized structure function $g_1(x)$, and may help to clarify
the discrepancies between the previous and recent data of $g_1$ at
$x\sim 10^{-2}$, and between the Regge extrapolation and the QCD fit for
$g_1$ at $x\to 0$ \cite{Le}.

It is well-known that the soft divergences from virtual and real 
radiative corrections to DIS cancel each other order by order. However,
in the derivation of the CCFM equation virtual gluons are summed to all
orders and grouped into the Sudakov and non-Sudakov form factors, while
real gluon contributions are evaluated only to lowest order, which leads
to the splitting function $P_{gg}$. To fulfill the soft cancellation,
the coupling constant ${\bar\alpha}_s$ in the exponent of the Sudakov
form factor $\Delta_S$ must be frozen, such that the lower bound $q$
of the variable $p^+$ can reach zero, {\it i.e.}, $\Delta_S\to 0$ as shown
in Eq.~(\ref{nons}). The infrared singularity from the factor $1/q^2$ in
the kernel of Eq.~(\ref{ccfm}) is thus suppressed. Otherwise,
a running ${\bar\alpha}_s({\bar\mu})$ will prohibit $q$ from being
below $\Lambda_{\rm QCD}$. However, the constant ${\bar\alpha}_s$ becomes
a parameter. With our approach, it is trivial to derive a new unified
evolution equation for the $\ln Q$ and $\ln(1/x)$ summations, in which both
virtual and real corrections are considered to lowest order \cite{LL}. 

At last, to improve the accuracy of the kernel to next-to-leading logarithms, we only 
need to evaluate the $O(\alpha_s^2)$ subdiagram. Such an evaluation can be 
performed in a straightforward way. The BFKL equation including the 
summation of the next-to-leading $\ln(1/x)$ will be published elsewhere.

\vskip 0.5cm
This work is supported by National Science Council of R.O.C. under the 
Grant No. NSC87-2112-M-006-018.

\newpage
\centerline{\large \bf Figure Captions}
\vskip 0.5cm

\noindent
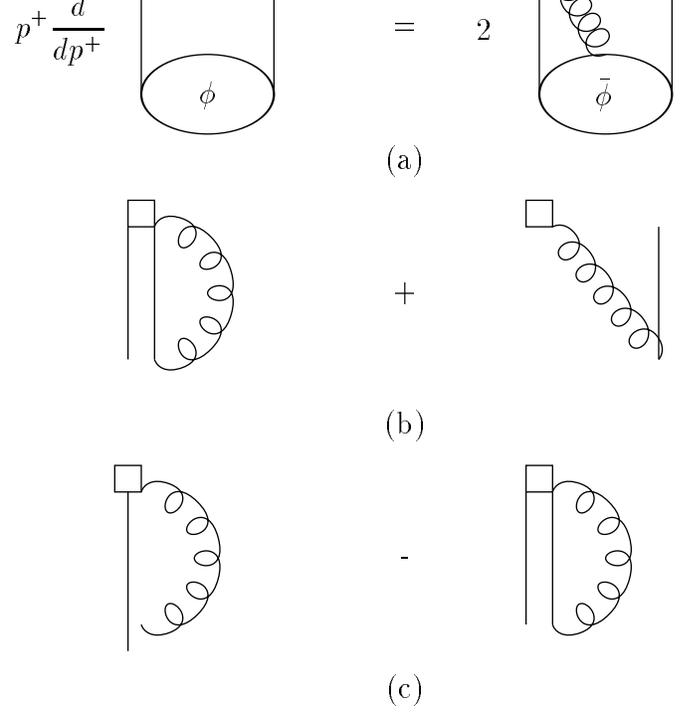
{\bf FIG. 1.} (a) The derivative $p^+d\phi/dp^+$ in the axial gauge.
(b) The soft structure and (c) the ultraviolet structure of the
$O(\alpha_s)$ subdiagram containing the special vertex.
\vskip 0.5cm

\noindent
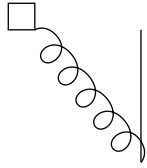
{\bf FIG. 2.} One of the diagrams that contributes to the finite part
of $P_{qq}$.
\vskip 0.5cm

\noindent
{\bf FIG. 3.} (a) The derivative $-xdF/dx$ in the axial gauge.
(b) The soft structure and (c) the ultraviolet structure of the
$O(\alpha_s)$ subdiagram containing the special vertex.
\vskip 0.5cm

\noindent
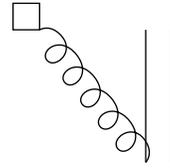
{\bf FIG. 4.} (a) The subdiagram containing the special vertex for the CCFM 
equation. (b) The subdiagram for the CCFM equation after resumming the 
double logarithms in $J$.

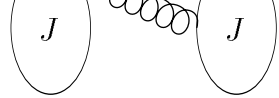
\end{document}